\documentclass[preprint]{aastex6}

\def\bfc{}

\usepackage{graphicx}
\usepackage{epsf}
\usepackage{natbib}

\newcommand{\simlt}{\lower.5ex\hbox{$\; \buildrel < \over \sim \;$}}

\providecommand{\sorthelp}[1]{}

\begin{document}

\title{Variations between Dust and Gas in the Diffuse Interstellar Medium. 3. Changes in Dust Properties}

\shorttitle{Dust Property Variations}

\author{William T. Reach}
\affiliation{Universities Space Research Association, MS 232-11, Moffett Field, CA 94035, USA}
\email{wreach@sofia.usra.edu}

\author{Jean-Philippe Bernard}
\affiliation{Universit\'e de Toulouse, Institut de Recherche en Astrophysique et Plan\'etologie, F-31028 Toulouse cedex 4, France}

\author{Thomas H. Jarrett}
\affiliation{Astrophysics, Cosmology and Gravity Centre, Astronomy Department, University of Cape Town, Private Bag X3, Rondebosch 7701, South Africa}

\author{Carl Heiles}
\affiliation{Astronomy Department, University of California, Berkeley, CA 94720, USA}

\begin{abstract}
We study infrared emission of
17 isolated, diffuse clouds with masses of order $10^2 M_\odot$, to test the hypothesis that
grain property variations cause the apparently low gas-to-dust ratios that have been measured
in those clouds. Maps of the clouds were constructed from {\it WISE} data and directly compared
to the maps of dust optical depth from {\it Planck}. The mid-infrared emission per unit dust optical
depth has a significant trend toward lower values at higher optical depths. The trend can be
quantitatively explained by extinction of starlight within the clouds. The relative amounts of PAH and
very small grains traced by {\it WISE}, compared to large grains tracked by {\it Planck}, are consistent
with being constant. The temperature of the large grains significantly decreases for clouds with larger dust 
optical depth; this trend is partially due to dust property variations but is primarily due to extinction of starlight.
We updated the prediction for molecular hydrogen column density, taking into account variations 
in dust properties, and find it can explain the observed dust optical depth per unit gas column density.
Thus the low gas-to-dust ratios in the clouds are most likely due to `dark gas' that is molecular hydrogen.
\end{abstract}

\keywords{
dust, ISM: abundances, ISM: atoms, ISM: clouds, ISM: general,  ISM: molecules
}

\section{Introduction}

The nature and content of interstellar clouds can be measured through various tracers including
the 21-cm line of atomic hydrogen, far-infrared emission of dust, microwave emission lines of CO, 
ultraviolet and optical absorption lines, extinction by dust, and $\gamma$-rays. 
These diverse tracers reveal material
with the range of physical conditions and geometrical observing conditions in which they apply,
leading to an overall understanding of the transition in the interstellar medium
between low-density ionized gas, diffuse atomic gas, translucent clouds, dark clouds, and giant molecular
clouds \citep{snow06}. The relative amount of atomic and molecular material in the diffuse and translucent clouds, where
H$_2$ molecules form on dust grain surfaces with column density limited by destruction from ultraviolet
photons in the interstellar radiation field \citep{hollenbach71}, remains uncertain at the factor of 2 level
because H$_2$ is not readily directly observed \citep{xfactorreview}.
We initiated a study of isolated, approximately degree-sized, approximately $10^2 M_\odot$, interstellar clouds that allow clear separation 
between the cloud and the unrelated diffuse interstellar medium and that have a range of brightnesses and locations across the sky.
We showed in Paper 1 \citep{reach15} that such clouds have a wide range of gas-to-dust ratios. To 
explain them, we asserted three hypotheses: 
\begin{enumerate}
\item[(1)] The amount of gas was underestimated due
to the presence of extensive molecular hydrogen not accounted for by the 21-cm observations of $N_{\rm H}$.
\item[(2)] The amount of gas was underestimated due to the 21-cm line being optically thick.
\item[(3)] The amount of dust was overestimated by the submm opacity because of variations in the dust properties.
\end{enumerate} 
In all cases we retain the ansatz that the {\it actual} gas-to-dust mass ratio remains constant, which 
is quite likely because there are no evident local sources of dust in the clouds nor mechanisms to separate the gas from dust.
Recent theoretical work supports that interstellar grains remain closely coupled to the gas even in molecular clouds (and even
neglecting magnetic fields), with only rare grains larger than 1 $\mu$m sized grains experiencing significant segregation \citep{tricco17}. 
We fully expect that all three hypotheses are active, and our goal is to measure the relative
importance of each.
Paper 1 showed a simple analytical calculation \citep[from][]{reach94}
for the formation of molecular
hydrogen and showed that hypothesis (1) could match 
the observations with no free parameters. In Paper 2 \citep{reach17}, we showed, from 21-cm absorption observations toward radio continuum sources behind the clouds, that an apparently low gas-to-dust ratio 
could not be fully explained by cold atomic gas, because 
there is not enough opacity measured in the 21-cm line. While there
is clearly some cold, atomic gas that explains a small part of the high specific opacity, 
Paper 2 effectively ruled out hypothesis (2). 
In this paper, we test hypothesis (3). 

The initial sample for comparison between far-infrared dust and
\ion{H}{1}  gas distribution defined in Paper 1;
that sample was restricted to the Arecibo declination range. We maintained that sample for Paper 2, which again
used Arecibo 21-cm observations (of absorption this time).
For this paper, we expand the sample to have a more diverse set of clouds for studying potential
dust property variations. The supplemental clouds are from \citet{hrk88}, where isolated, degree-sized
clouds were identified from inspection of the {\it IRAS} 100 $\mu$m images and the all-sky \ion{H}{1} images at half-degree
resolution. To further diversify the sample, we added 
DIR015+54, which has a distinctively warmer dust temperature than other high-latitude clouds,
despite having
no internal heating source \citep{reach98}.

For all clouds, we used maps of the dust opacities at $5'$ resolution from the {\it Planck} 
\citep{tauber2010a} survey, specifically the `thermal dust' foreground separation
\citep{planck2013viii}, to compare to mid-infrared emission maps from the {\it Wide Field Infrared Survey
Explorer (WISE)} all-sky survey \citep{wright10}, specifically the {\it WISE} Image Atlas served through
the Infrared Science Archive.

\section{Properties of Large Grains \label{bgsec}}

Models for interstellar grain optical properties and size distribution have been developed to simultaneously
fit the UV/optical/near-infrared extinction and the thermal emission from diffuse interstellar lines of sight.
Recent models include that of \citet{lidraine} that comprises solid particles of `astronomical silicate' and
carbonaceous material supplemented by polycyclic aromatic hydrocarbons (PAHs),
and that of \citet{jones13} that comprises silicate and carbonaceous particles with mantles of varying depth.
When tuned to match a single environment, the models can match the observations and provide largely self-consistent
assessments of abundances of elements that compose the grains, together with predicted observables at a wide range of wavelengths. 
\def\extra{If the dust properties are fixed, then one can use the models to calculate the emission and
extinction in environments other than where the models were fitted. Thus, the dust models could in 
principle be used
in molecular clouds or protoplanetary disks, in supernova remnants or near active galactic nuclei.}

To utilize dust models in environments other than where they were fitted requires assuming dust
properties remain unchanged. This can be tested observationally.
Some of the first submm observations of interstellar clouds by {\it PRONAOS} showed that the apparent dust emissivity
is significantly (factor-of-three) enhanced in molecular clouds compared to the diffuse interstellar 
material \citep{stepnik03}. 
In the well-characterized Taurus molecular cloud, far-infrared observations with {\it Spitzer} 
and submm observations with {\it Planck} showed
a factor-of-two increase in specific opacity in the molecular regions compared to 
atomic regions \citep[][]{flagey09,planckXXVtaurus}.
\citet{planckXXIVdust} studied 14 large high-latitude fields, comparing dust emission to dedicated observations of
the 21-cm line with the Green Bank Telescope, and they found significant variation. 
Variations in dust properties were conclusively demonstrated by \citet{planckXVIIdust}, who showed that even
in low column density, high-latitude lines of sight, where there are no variations of radiation field due to extinction or local heating sources, there is a clear trend of specific dust opacity versus temperature.
The quantity that is
directly observable is the specific opacity, which is the optical depth per unit gas column density,
\begin{equation}
\sigma_{353} \equiv \tau_{353} / N_{\rm H~I} 
\end{equation}
where $\tau_{353}$ is the optical depth at a convenient reference frequency of 353 GHz
(equivalent to a wavelength of 850 $\mu$m), and $N_{\rm H}$ is the gas column density (from \ion{H}{1} 21-cm line observations and,
where available, CO observations converted to H$_2$ column density and doubled). The value of $\sigma_{353}$ is derived by correlation between 
dust and gas observations scaled to identical spacial scales.
To convert the specific opacity to other frequencies requires scaling by the frequency,
which in the far-infrared and submm is normally expressed as a power law
\begin{equation}
\sigma_\nu = \sigma_{353} \left(\frac{\nu}{353{\rm ~GHz}}\right)^\beta,
\end{equation}
even though in reality materials could have a more complex frequency dependence of absorption cross-section. The dust models mentioned above were developed to match the specific opacity in the diffuse interstellar medium;
the model of \citet{lidraine} matches the observed, typical specific opacity with a gas-to-dust mass
ratio of 124 and usage of the elements in accord with cosmic abundances and observed depletions of
those elements into grains.

Now that the {\it Planck} results have been studied by multiple groups and validated, 
the interstellar dust models will need to be updated. The empirical trends that will drive
the most significant updates include the trend of specific opacity with dust temperature 
\citep{planckXVIIdust,planckXXIVdust} and the discrepancy in the ratio of far-infrared to visible
optical depth \citep{planckXXIX}. The former directly affects the inferred amount of gas for
a given amount of dust, and the latter would affect the temperatures of the grains, which absorb
visible-UV light and emit in the far-infrared. 
A first systematic evaluation of the models by \citet{fanciullo15} already shows that the 
observational results cannot be explained by a uniform dust properties, even with variable
illumination by the interstellar radiation field.

In Paper 1, we measured the specific opacities of a set of high-latitude clouds with high-resolution
21-cm observations from Arecibo. For convenience, we converted the specific opacities into gas-to-dust ratios
using the \citet{lidraine} model:
\begin{equation}
\label{eq:gddef}
[G/D] = \frac{9.7\times 10^{-25} {\rm ~cm}^{2} \rm{ ~H}^{-1}}{ \sigma_{353}}.
\end{equation}
In this paper we prefer the model-independent observable $\sigma_{353}$, but as equation~\ref{eq:gddef}
shows, the gas-to-dust ratios are readily converted back to $\sigma_{353}$.

\begin{figure}
\plotone{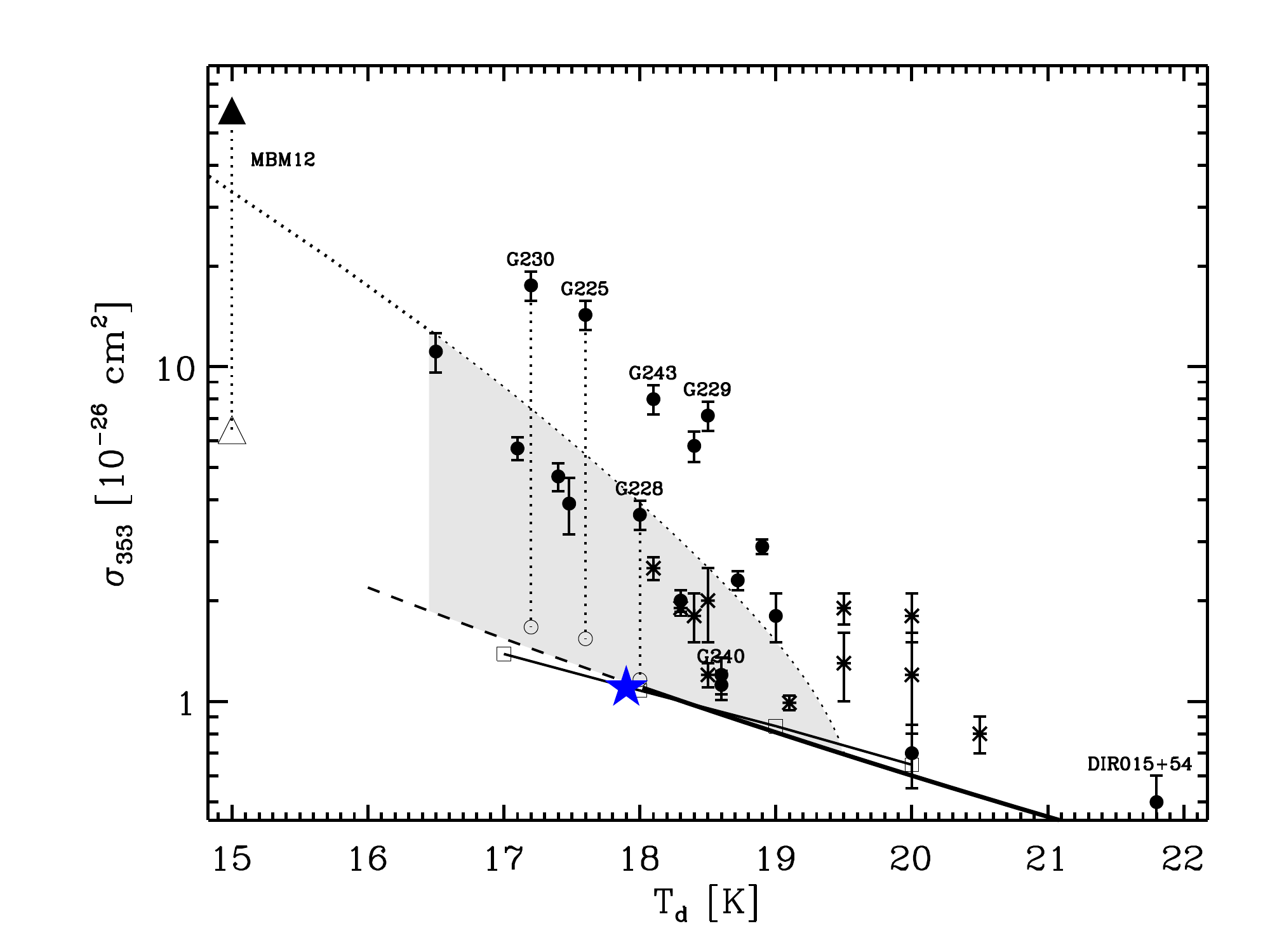}
\caption{The specific opacity, dust opacity divided by the 21-cm H~I column density (solid symbols), versus dust temperature
for the diffuse clouds in this study (filled circles).
The large blue star is the high-latitude sky average derived by \citet{boulanger96}
using low-resolution {\it COBE} and H~I data, scaled to 353 GHz.
The triangles are the high-latitude molecular cloud MBM~12: the dust temperature is from {\it Planck}; the dust optical
depth is based on the {\it Spitzer}/MIPS 160 $\mu$m brightness and {\it Planck} temperature with emissivity scaling to 353 GHz; and the
atomic gas column density is toward the core \citep[{\it upper point};][]{peek11}.
For clouds with CO detections from \citet{reach94} or from \citet{pound90}, an open symbol shows the specific opacity calculated using the total (atomic and molecular) gas;
a dotted line connects $\sigma_{353}$ determined with atomic and total gas. 
The thick line is the trend observed in low column density lines of sight from \citet{planckXVIIdust}, and the dashed line
is a continuation of that trend to lower temperatures.
The thin, solid line with squares is the trend observed for diffuse regions from \citet{planckXXIVdust}. 
The thick, dotted line shows the model for H$_2$ formation based on Paper 1 but updated to the dust property variations.
The shaded region highlights the region where molecules significantly contribute to the column density and can explain the
observed $\sigma_{353}$.
\label{dgvarfig}}
\end{figure}

Figure~\ref{dgvarfig} shows the specific opacity for the clouds versus their
dust temperature.
The specific opacities are adapted from the gas-to-dust ratios in Paper 1 as
per equation \ref{eq:gddef}.
The dust temperatures were measured for all clouds
by fitting a modified blackbody to the surface brightness from {\it IRAS} and {\it Planck} at frequencies 3000 to 353 GHz (wavelengths 100 to 850 $\mu$m) as described in \citet{planckAllSkyDust}.
The main results from Papers 1 and 2 were measurement of the change in apparent gas-to-dust ratio and demonstration  that 
 coherent regions in interstellar clouds have either either `dark' gas or dust property evolution toward higher specific opacity at low temperatures.

We now compare the cloud observations to 
recently-published trends in
specific opacity due to grain property variations.
Figure~\ref{dgvarfig} shows that the apparent trends toward higher specific opacity at lower temperature are
in the same general sense for our cloud sample (symbols) as the trend for the diffuse ISM (straight lines). 
But the apparent trend is significantly more pronounced for our isolated cloud sample than for the diffuse ISM: clouds at low temperature have significantly higher emissivity, as seen by the symbols
appearing far above the straight lines. 
The diffuse ISM does not contain gas colder than approximately 18 K, so to make a direct comparison we extrapolated the dust property variation trend to lower
temperatures (straight dashed line). 
Both of these lines show the trend predicted for regions that have the same total power emission but have dust properties that vary with temperature; such regions would have
constant $\sigma T^{4+\beta}$. The approximate power-law fit to the diffuse ISM for $\beta=1.8$ is
\begin{equation}
\label{eq:sigism}
\sigma_{353}=1.1\times 10^{-26} \left(\frac{18}{T_d}\right)^{5.8} \,\,\,{\rm cm}^{2}.
\end{equation}
It is evident the coldest clouds in our sample have specific opacities a factor of few higher
than the extrapolated trend of dust properties derived in the diffuse interstellar medium.
We therefore conclude that the recently-observed variations in dust properties as a function of temperature 
explain {\it some but not all} of the apparently high specific opacity 
in isolated high-latitude clouds.

In Paper 1, we demonstrated that a simple analytical model for H$_2$ formation can explain
the apparently higher specific opacity in colder clouds. 
We can now update this calculation (which was our Hypothesis 1 to explain the enhanced
specific opacity) taking into account the independently observed trends in dust
properties in the diffuse ISM (which could be a lower limit for our Hypothesis 3). 
The H$_2$ formation model in Paper 1 approximately matched the
observations, with essentially no free parameters. Figure~\ref{dgvarfig} shows the updated predictions,
where we simply took the model from Paper 1 and scaled the dust specific opacity at each temperature to match equation~\ref{eq:sigism}. We did not change the H$_2$ formation rate
on grain surfaces as a function of dust properties, though we anticipate such an effect theoretically; a prediction for this effect awaits future laboratory-result-motivated 
theoretical work.
From Figure~\ref{dgvarfig}, the model for H$_2$ formation remains consistent with the 
observations, with the variation in dust properties being a smaller effect than
formation of molecular hydrogen.

While the apparent `dark gas'  can be 
plausibly explained by formation of H$_2$ (Hypothesis 1), we must remain open-minded to the possibility that dust property variations could explain even more of the opacity changes, if those dust property variations
are {\it much more extreme} in our clouds than observed in the diffuse ISM---for example by  runaway dust coagulation. 
In that case, we expect the PAH and very small grains could be coagulated onto the larger grains in the cloud cores. This effect can be directly tested using mid-infrared observations, which is the subject
of the next section.


\section{Abundance of Small Grains \label{pahsec}}

\begin{figure}
\plotone{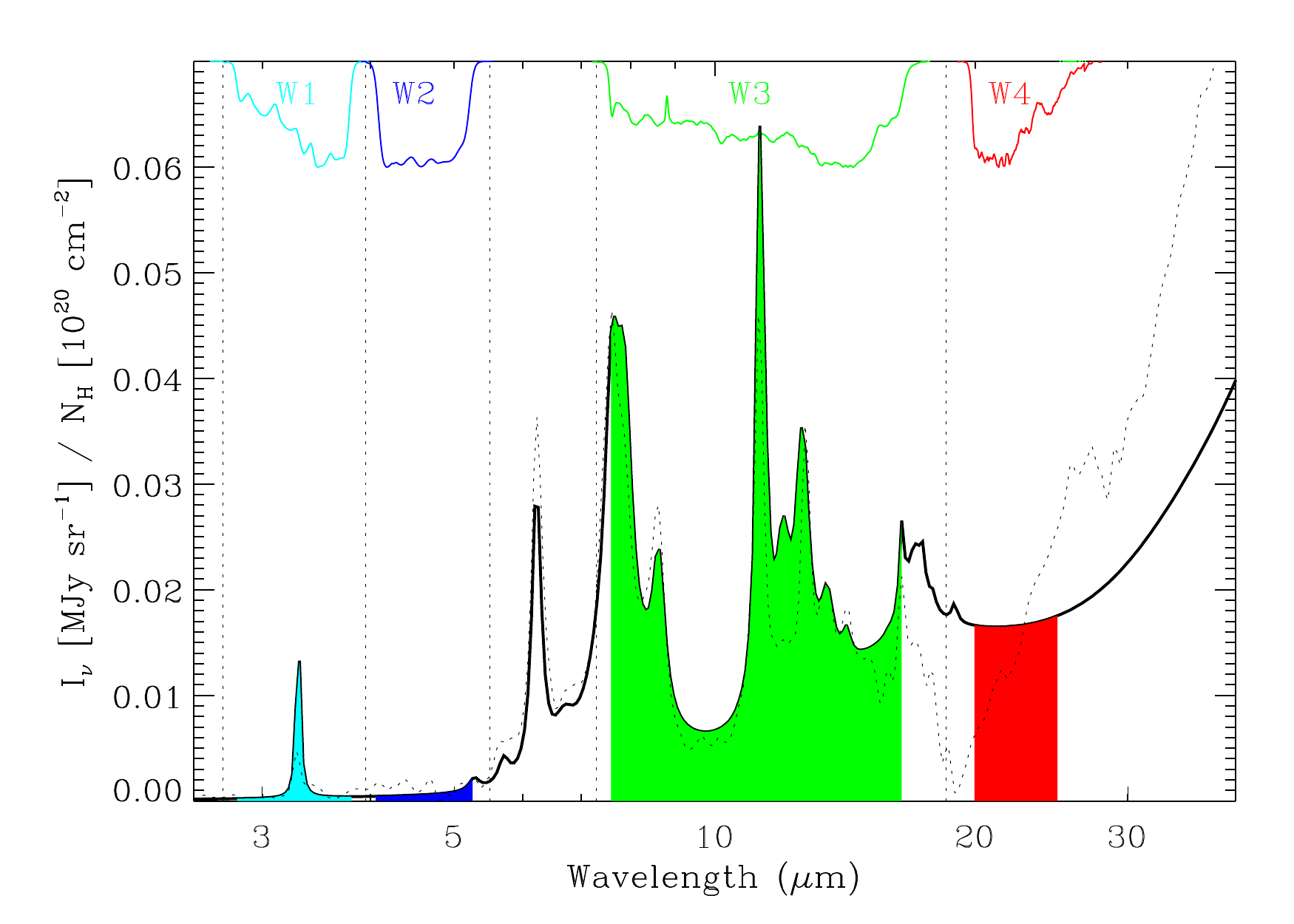}
\caption{Emissivity of the diffuse interstellar medium \citep{lidraine} (thick curve), together with the {\it WISE} relative spectral response functions (shown inverted at the top of the figure for each of the 4 channels). The regions with relative spectral response greater than 10\% of the maximum for each channel are colored accordingly (channel 1 in cyan, channel 2 in blue, channel 3 in green, channel 4 in red). The dotted curve shows the {\it ISO}/SWS
spectrum of NGC 7023, with fluxes in Jy divided by 3000. 
\label{wisespec}}
\end{figure}

The mid-infrared surface brightness of interstellar clouds is due to non-equilibrium emission from transiently
heated particles, as explained e.g. by \citet{da85}. At wavelengths shorter than 18 $\mu$m, interstellar emission
is predominantly from macromolecules including polycyclic automatic hydrocarbons (PAHs). Figure~\ref{wisespec}
shows a nominal interstellar spectrum from a dust model tuned to match the observations
of the diffuse interstellar emission, as well as the spectrum of reflection nebula NGC 7023, taken from the
{\it Infrared Space Observatory} archive.
The PAH features at 3.3, 
6.2, 7.7, 8.6, 11.3, 12.6, and 16--18 $\mu$m are readily evident, as is the rapid rise toward longer wavelengths due to transiently heated, very small grains (VSGs). 
To assess the variations in PAH and VSG abundances in interstellar clouds, we will use data from the
{\it Wide Field Infrared Explorer} ({\it WISE}), which surveyed the entire sky in four broad wavebands. 
Figure~\ref{wisespec} shows which part of the interstellar spectrum is sampled by each of the bands.
The W1 band contains the 3.3 $\mu$m PAH feature; W2 has relatively little emission; W3 contains multiple PAH features; 
and W4 is dominated by steeply rising  emission from very small grains (VSG). 
Because stars and galaxies are so bright and densely populate the sky at the shorter wavelengths,
we concentrate on the W3 and W4 bands, which address our primary concern on grain abundance. 
Both W1 and W2 also contain contributions from `cloud shine' due to scattering of
the interstellar radiation field by large particles \citep{foster06,andersen13}. Given the difficulty of
separating the diffuse emission from stars and galaxies in W1 together with the need to distinguish `cloud shine'
from the PAH emission, we do not utilize W1 in this study despite the presence of the 3.28 $\mu$m feature in that band.
Figure Set 3 shows the {\it WISE} band 3 images 
of each cloud, and the clouds are readily evident as being well traced by {\it WISE}. The relatively high
angular resolution of {\it WISE} reveals new details, such as intricate, filamentary structures, which we intend to study in future, detailed studies of individual clouds. 

\def\extra{
\begin{figure}
\includegraphics[width=6.6in]{G236p39_wise234.png}
\caption{{\it WISE} image of the cloud G236+39.
This RGB composite image is made with blue being
the W2 band, dominated by stars, green being the W3 band, dominated by PAH, and red being the W4 band, dominated by VSG. The image size is $4.5^\circ\times 4.5^\circ$, oriented with celestial North upward.
\label{G236p39wise234}}
\end{figure}
}

In this paper we concentrate on the comparison between {\it WISE} and {\it Planck}, with the latter having
an angular resolution of $5^\prime$. 
The {\it WISE} and {\it Planck} images are compared in detail in
Figure Set 3.
It is evident that {\it WISE} and {\it Planck}
reveal  similar cloud morphologies and dynamic ranges of infrared surface brightness and optical depth, making
them ideal for studying dust property variations within clouds and from cloud to cloud. The need for 
expanding the survey beyond that used in Paper 1 is also evident. The clouds from Paper 1 included some in
the ecliptic plane (e.g. G254+63), for which the variations in the mid-infrared Solar System dust emission are as large
as those of the interstellar clouds. All such clouds can be discerned in the {\it WISE} images, 
though their surface brightness is measured only 
approximately. The clouds we added to the sample from HRK, on the other hand, were generally brighter and more cleanly separated from unrelated structures.

For each cloud, we extracted the surface brightness along a slice that extends from well outside, through the central concentration, and well out the other side of the cloud. A polynomial baseline was then fitted to the regions outside the cloud in order to remove unrelated interstellar emission. In some cases, this extended emission may indeed be related to the cloud as a diffuse envelope, but we nonetheless remove it in order to generate an identically-determined, differential 
measurement of the cloud brightness. 
Figure Set 3 shows the surface brightness profiles in
dust optical depth, $\tau_{353}$, and {\it WISE} band 3 (W3) and band 4 (W4) brightness. The fine details of the clouds
are not considered in this study; differences seen in the brightness profiles are at least partially due to
the difference in angular resolution between {\it WISE} and {\it Planck}.

\begin{figure}
\figurenum{3}
\plotone{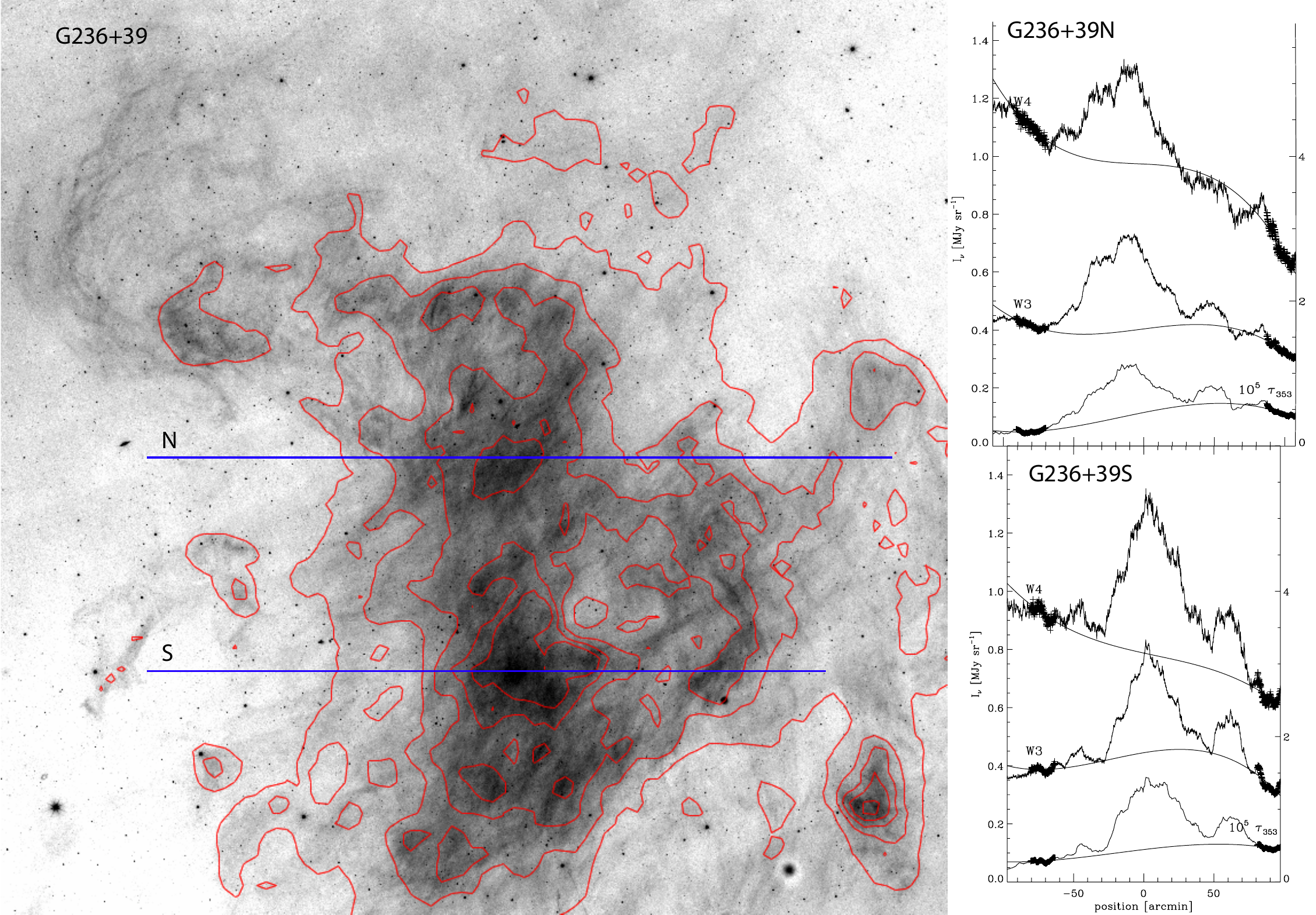}
\caption{{\it WISE} band 3 images of each cloud, with contours of dust optical depth
from Planck overlaid in red. 
The printed article the example of G236+39.
The complete figure set (13 images) is available in the online journal.
The greyscale ranges from 0 to 0.63 MJy~sr$^{-1}$ (relative to an arbitrary background level). Contour values are at $\tau_{353}=(3.4, 5.6, 7.8, 10, 12)\times 10^{-6}$. The image size is $4.5^\circ\times 4.5^\circ$, oriented with celestial North upward. The purple line shows the location of the surface brightness profile, which is displayed in the right-hand panel. The dust optical depth ($\tau_{353}$) units are on the right-hand vertical scale, and the mid-infrared surface brightness (W3 and W4) units are on the left-hand vertical scale of the spatial profile plot.
}
\end{figure}

\clearpage

For each cloud, we correlated the baseline-subtracted {\it WISE} surface brightness profile
and {\it Planck} optical depth profile. Table~\ref{fittab} compiles the results.
In all cases the {\it WISE}
band 4 and 3 brightness correlate very tightly with each other, and slopes,
$d$W4/$d$W3 in Table~\ref{fittab},
have statistical uncertainty less than 1\%. The total uncertainty in $d$W4/$d$W3 is set
by the {\it WISE} calibration accuracy of 6\% at 22 $\mu$m.
The slope of the correlations between {\it WISE} surface brightness and large-grain optical depth is the emissivity,
\begin{equation}
\epsilon(\lambda) \equiv \frac{I_\nu}{\tau_{353}};
\end{equation}
we use MJy~sr units for the {\it WISE} surface brightnesses at 12 $\mu$m and 22 $\mu$m to obtain
$\epsilon(12)$ and $\epsilon(22)$, respectively.
As should be expected $d$W4/$d$W3 and $\epsilon(22)/\epsilon(12)$ are similar, with values
in the range 0.8--1.5. These quantities were derived by
completely different methods, so on a cloud-by-cloud basis there is a standard deviation between
the methods of 25\%. For the cloud color we recommend the more-precise $d$W4/$d$W3.

\begin{figure}
\plotone{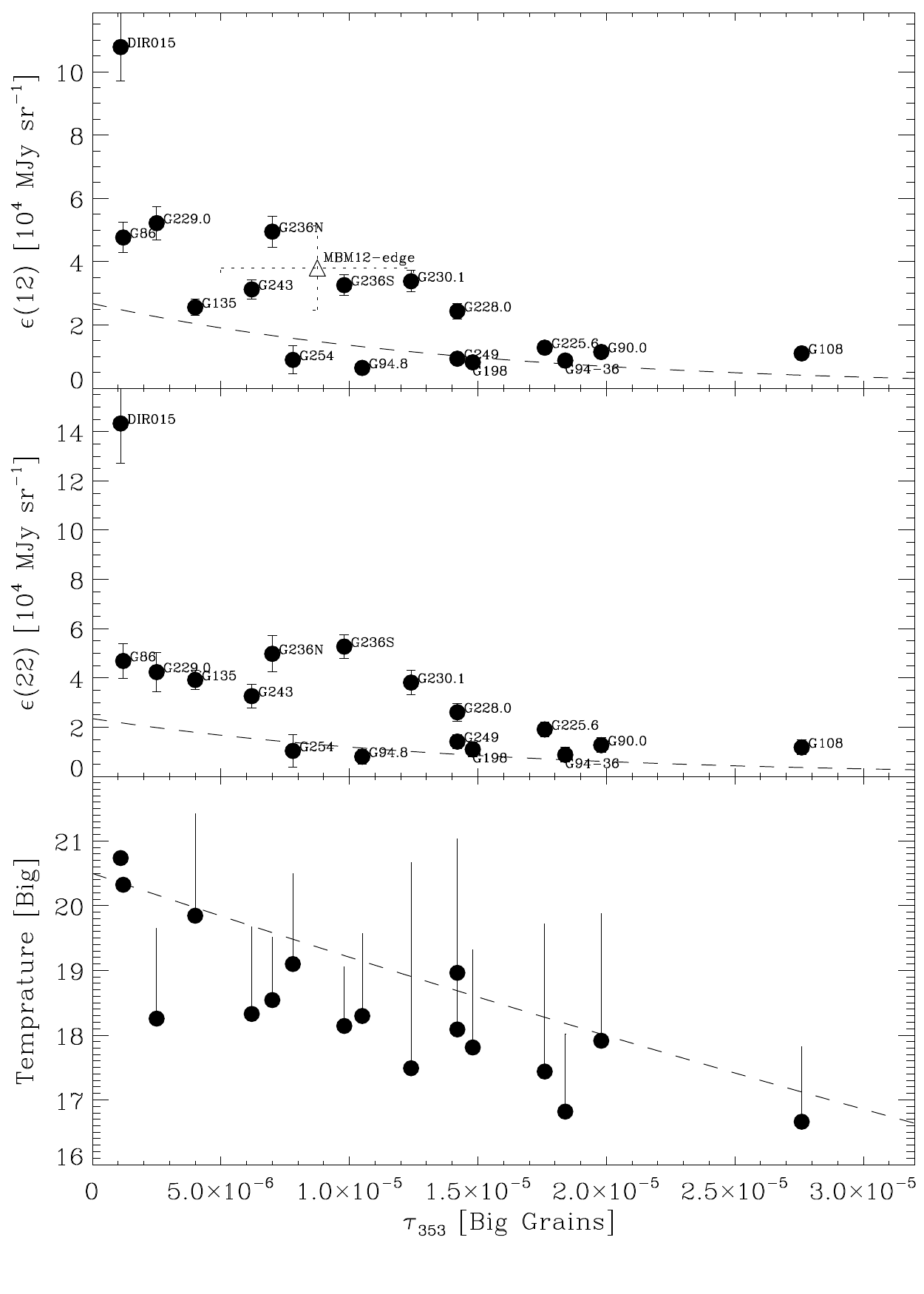}
\caption{Comparisons between PAH, VSG, and Big Grains derived from the {\it WISE} and {\it Planck} observations.
The top panel and middle panels show the {\it WISE} band W3 (12 $\mu$m) and W4 (22 $\mu$m) brightness per unit 
{\it Planck} optical depth, measuring the ratio of PAH and VSG per unit big grains, plotted against the {\it Planck}
optical depth. There is a general trend such that there is relatively less mid-infrared emission per unit 
dust column density where there is
more higher dust column density (potentially due to a combination of shielding of the PAH from the ISRF combined with sticking 
of the PAH onto the Big Grains). 
The lower panel shows the large grain dust temperature as a function of {\it Planck} optical depth. The filled
circle is for the cloud center, and the vertical bar extends upward to the temperature just outside the cloud.
Clouds are labeled in the top and middle panels by the initial portion of their names from Table~\ref{fittab}.
Dashed lines are empirical models discussed in the text.
\label{w3tau}}
\end{figure}

\def\tnm{\tablenotemark}
\begin{deluxetable}{lccccccc}
\tabletypesize{\scriptsize} 
\tablecolumns{7}
\tablecaption{Mid-Infrared Emissivities for Diffuse Interstellar Cloud\label{fittab}} 
\tablehead{
\colhead{Cloud} & \colhead{$N_{\rm H}$\tablenotemark{a}} & \colhead{$T_{\rm dust}$} & \colhead{$n_{\rm H}$\tablenotemark{b}} & \colhead{$d$W4/$d$W3} &  \colhead{$\epsilon(12)$} & \colhead{$\epsilon(22)$} & Figure\\
                  &    (10$^{20}$ cm$^{-2}$)            & {(K)}       & (cm$^{-3}$)         & & {($10^{-4}$ MJy~sr$^{-1}$)} & {($10^{-4}$MJy~sr$^{-1}$) } 
}
\startdata
           DIR015+54 &  1.1 & 20.6 & 22 & 1.31 & $ 10.8 \pm 1.1$ & $ 14.4 \pm 1.6$ & 3.1 \\
              G86+59 &  1.2 & 20.3 & 23 & 0.70 & $  4.8 \pm 0.5$ & $  4.7 \pm 0.7$ & 3.2 \\
         G90.0+38.8  &  20 & 17.9 & 1000 & 1.10 & $  1.1 \pm 0.4$ & $  1.3 \pm 0.6$ & 3.3 \\
         G94.8+37.6  &  10 & 18.3 & 550 & 1.29 & $  0.6 \pm 0.4$ & $  0.8 \pm 0.6$ & 3.3 \\
              G94-36 &  21 & 16.8 & 160 & 1.45 & $  0.5 \pm 0.4$ & $  0.5 \pm 0.6$ & 3.4 \\
       G108-53       &  27 & 16.7 & 600 & 0.92 & $  1.1 \pm 0.4$ & $  1.2 \pm 0.6$ & 3.5 \\
          G135+54    &  4.0 & 19.8 & 86 & 1.64 & $  2.5 \pm 0.4$ & $  3.9 \pm 0.6$ & 3.6 \\
             G198+32 &  16 & 17.8 & 200 & 0.79 & $  0.8 \pm 0.4$ & $  1.0 \pm 0.6$ & 3.7 \\
      G225.6-66.4    &  18 & 17.4 & 1900 & 1.50 & $  1.3 \pm 0.4$ & $  1.9 \pm 0.6$ & 3.8 \\
      G229.0-66.1    &  2.5 & 18.3 & 430 & 0.76 & $  5.2 \pm 0.5$ & $  4.2 \pm 0.8$ & 3.8 \\
      G230.1-28.4    &  12 & 17.5 & 550 & 1.09 & $  3.4 \pm 0.4$ & $  3.8 \pm 0.6$ & 3.9 \\
      G228.0-28.6    &  14 & 18.1 & 390 & 1.18 & $  2.4 \pm 0.4$ & $  2.6 \pm 0.6$ & 3.9 \\
      G236+39S       &  6.4 & 18.1 & 240 & 1.12 & $  3.4 \pm 0.4$ & $  4.9 \pm 0.6$ & 3.10 \\
      G236+39N       &  6.6 & 18.7 & 160 & 1.29 & $  4.7 \pm 0.5$ & $  4.6 \pm 0.7$ & 3.10 \\
          G243-66    &  6.2 & 18.3 & 130 & 1.06 & $  3.1 \pm 0.4$ & $  3.3 \pm 0.6$ & 3.11 \\
             G249+73 &  15 & 19.0 & 200 & 1.90 & $  0.9 \pm 0.4$ & $  1.4 \pm 0.6$ & 3.12 \\
             G254+63 &  7.9 & 19.1 & 160 & 1.41 & $  0.9 \pm 0.4$ & $  1.0 \pm 0.6$ & 3.13 \\
             MBM 12  &  130 & 14.2 & 2100 & ... & $<0.030$ to 0.5 & ... & \ref{mbm12image} \\
 \enddata
\tablenotetext{a}{Column density of the cloud, based on the {\it Planck} optical depth, and $\sigma_{353}=10^{-26}$ cm$^{-2}$ H$^{-1}$, and distance 100 pc}
\tablenotetext{b}{Estimated average gas density in the cloud, assuming line-of-sight depth is the same as the 
sky-plane extent, and a distance of 100 pc for each cloud.}
\end{deluxetable}

\clearpage

{\bfc

\subsection{Small grains in a dark cloud: MBM 12\label{mbm12sec}}

\begin{figure}
\epsscale{1}
\plotone{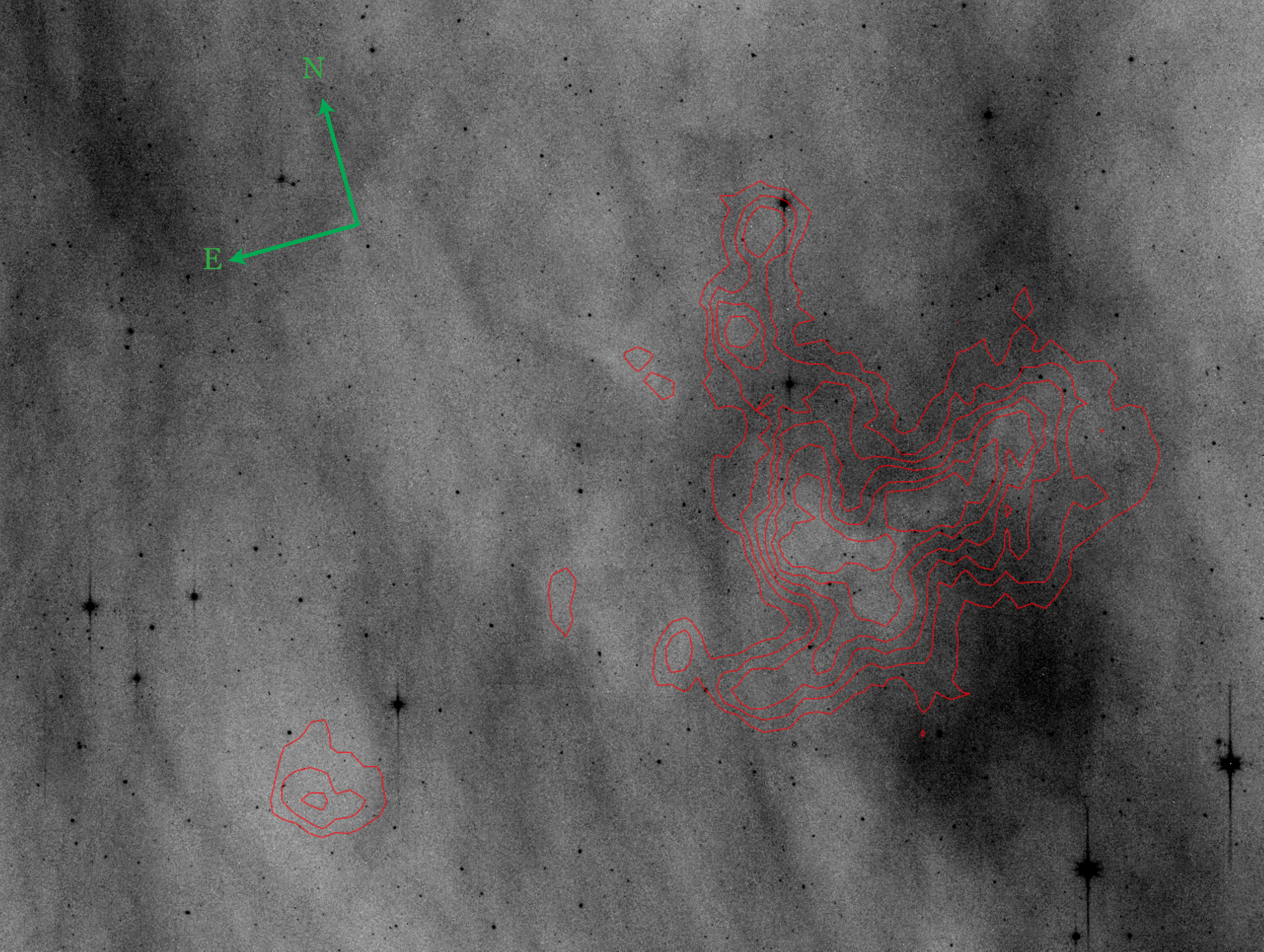}
\caption{Greyscale image of MBM 12 in the {\it Spitzer}/IRAC 8 $\mu$m band, with color scale range from 0 to 0.62 
MJy~sr$^{-1}$. Overlaid are red contours of the {\it Spitzer}/MIPS 160 $\mu$m brightness, at levels 
50, 61, 72, 83, 94, and 105 MJy~sr$^{-1}$. The image is centered on J2000 coordinates 02:56:40.4 +19:31:38, 
oriented with North and East as indicated by the vectors in the upper left. The lengths of the vectors are $5'$,
and the total image size is $51'\times 40'$. The 160 $\mu$m surface brightness increases gradually toward a peak
in the cloud core. In contrast, the 8 $\mu$m emission is bright at the cloud edges and drops to zero at the cloud
peak.
\label{mbm12image}}
\end{figure}

\begin{figure}
\plotone{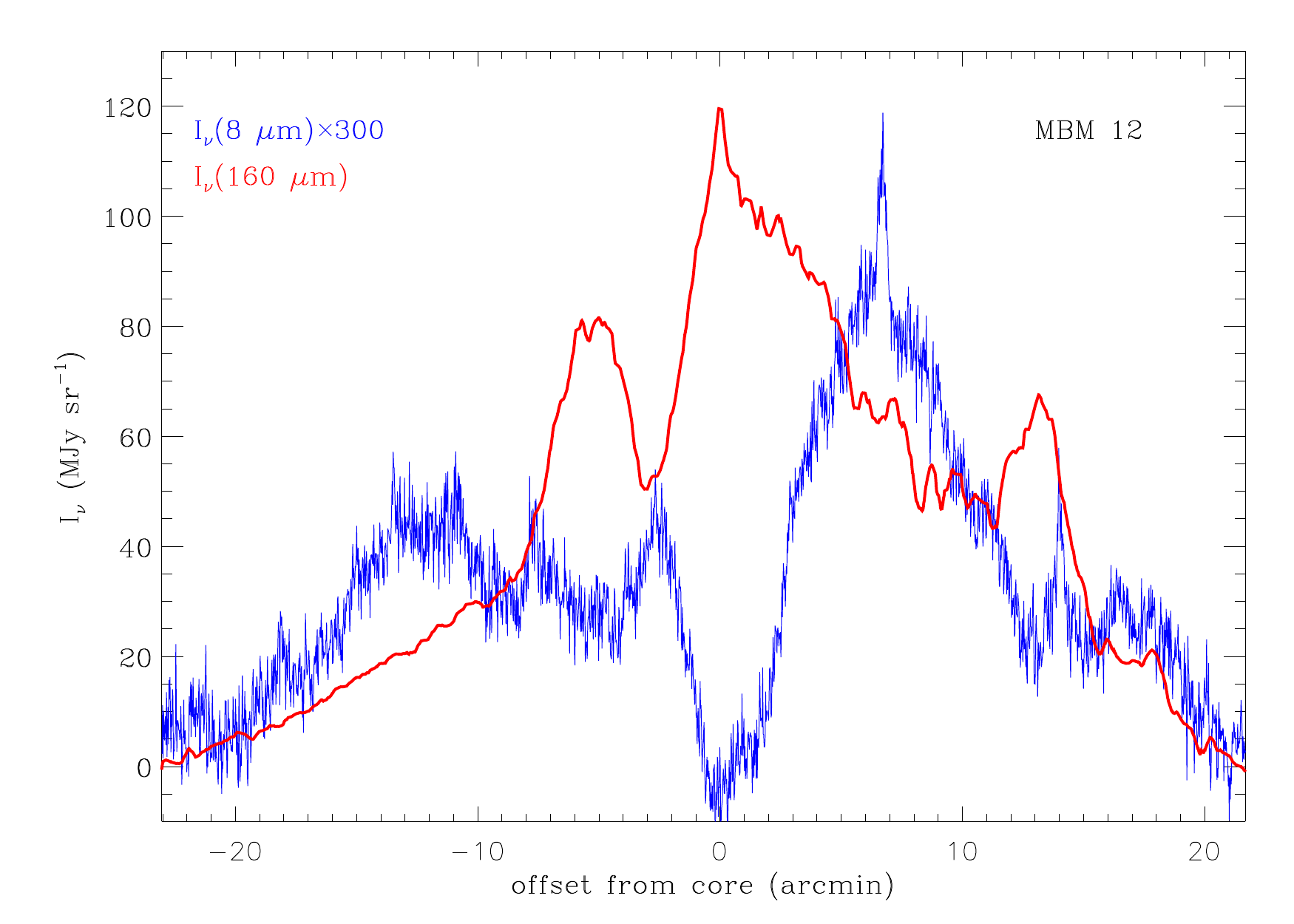}
\caption{Surface brightness profile of MBM 12 at 160 $\mu$m and 8 $\mu$m. The profile cuts vertically through the
region shown in Fig.~\ref{mbm12image}. At the cloud core, 160 $\mu$m emission peaks, while the
8 $\mu$m emission drops to zero. Two additional 160 $\mu$m peaks are present, and both are accompanied
by 8 $\mu$m decreases. 
\label{mbm12prof}}
\end{figure}

To extend the range of interstellar environments, and to enable comparison to studies large and small grain properties
in molecular clouds, 
we add a dark cloud that has high-quality mid- and far-infrared observations from {\it Spitzer}.
MBM 12 was discovered in the Palomar plates and confirmed as molecular
using mm-wave CO emission \citep{mbmfirst}. This cloud was observed by L. Magnani using {\it Spitzer}
on 2007 Sep 11 using the InfraRed Array Camera \citep[IRAC;][]{fazioirac} and 2007 Sep 23
using the Multiband Infrared Photometer for Spitzer \citep[MIPS;][]{rieke04}.
The 8 $\mu$m and 160 $\mu$m images were obtained from the {\it Spitzer} Heritage Archive.
Figure~\ref{mbm12image} compares these images toward the MBM 12 cloud core. The anti-correspondence between the
mid and far infrared images is striking. 
While the 160 $\mu$m image is relatively straightforward, with a general rise toward the cloud core
(including a small number of broad peaks) where the molecular cores are found in CO images
\citep{pound90}, the 8 $\mu$m image is filamentary and dispersed with the cloud core 
evident as a `shadow' where there is no emission.

To measure the dust emission properties quantitatively, and convert to a common scale with our {\it WISE} and {\it Planck} study, 
we use the spatial profile shown in Figure~\ref{mbm12prof}. Three distinct portions of the cloud can be identified.
The cloud `edges' are the low-column density regions near the edge of the image but with significant mid and far infrared emission.
The `halos' are the regions with the brightest 8 $\mu$m emission.
The `core' is the center of the cloud, which corresponds in 160 $\mu$m emission and other column density 
tracers including CO.
In each portion we measure the typical 160 $\mu$m surface brightness and ratio of 8 $\mu$m to 160 $\mu$m brightness.
For the cloud core, there is no evident emission so we measure the upper limit. These values are reported in the first columns
of Table~\ref{mbm12tab}. 
To compare to the {\it WISE} and {\it Planck} wavelengths used for the other clouds, we convert units as follows.
The 160 $\mu$m optical depth is determined by dividing the observed brightness by the Planck function evaluated at
dust temperature estimated from the {\it Planck} images; then the 353 GHz optical depth is determined using a dust
emissivity index of 1.75. The dust temperature and inferred optical depth are reported in Table~\ref{mbm12tab}.
The IRAC brightness was converted to equivalent {\it WISE} 12 $\mu$m surface brightness using a nominal ISM 
model emission spectrum \citep{lidraine}. The resulting emissivity is in the last column of Table~\ref{mbm12tab}.

Comparing to the translucent clouds that form our cloud sample, it is clear that the dark cloud MBM~12 is distinct. 
First, the far-infrared surface brightness and inferred optical depth of MBM~12 is significantly higher than our clouds.
Only the cloud edges overlap with our cloud sample, and even these edges only overlap with the brightest parts of our sample
clouds. In terms of mid-infrared emissivity, MBM~12 is low throughout. The cloud halos have mid-infrared
emissivities that overlap with
the weakest our sample clouds. The cloud core is lower by 2 orders of magnitude.

\begin{deluxetable}{lccccc}
\tablecolumns{5}
\tablecaption{Mid-Infrared Emissivities for portions of MBM 12\label{mbm12tab}} 
\tablehead{
\colhead{Portion} & \colhead{$I_{160}$} & \colhead{$I_8/I_{160}$} & \colhead{$T_{\rm dust}$} & \colhead{$10^5 \tau_{353}$} & \colhead{$\epsilon(12)$}\\
                  & (MJy~sr$^{-1}$)     &                       & {(K)}                   &                        & ($10^4$ MJy~sr$^{-1}$)
}
\startdata
Edge   &  15. & 0.0033 & 17.0 & 1.6 & 0.17\\
Halo   &  35. & 0.014 & 15.6 & 6.2 & 0.46 \\
Core   & 120. & $<0.00015$ & 14.2 & 28  & $<0.003$\\
 \enddata
\end{deluxetable}

}

\section{Discussion}

{\bfc
Having compiled the measured mid and far infrared brightness ratios for the clouds, we now discuss what are
the systematic trends and implications for dust property evolution in clouds.}
For context, a plausible scenario for explaining variations in
dust properties is an evolution of the mantles of the grains \citep{jones13}. The dust evolution would
include coagulation of carbonaceous
material onto larger grains in colder regions as one process, and the opposite effects
of fragmentation and mantle evaporation. 
\citet{ysard15} showed that the range of specific dust opacity could be explained by changes in dust mantle properties, in particular carbonaceous material accumulating onto larger grains and adding an extra 60 ppm of carbon
in solid form.
\citet{kohler15} showed that the timescale for VSG to accumulate onto big grains was less than $10^6$ yr, and they
suggest the trends in dust properties are due to the local gas density, which sets timescales and material
availability for accretion and coagulation.

\def\extra{
The driving forces of these processes could conceivably be
a dynamic balance between cloud-cloud collisions, supernova shock waves, and gravitational collapse.
}


\subsection{Comparison between grains in isolated clouds and the diffuse ISM}

The observational results clearly show a decrease in the emissivity of mid-infrared emission at higher column densities. 
First, we compare the cloud brightness to the diffuse, low-column-density ISM using the high-latitude sky
averages measured by \citet{arendt98} found from {\it COBE}/DIRBE data: 
$I_\nu/N({\rm H})=0.029$ and 0.030 MJy~sr$^{-1}$ per 10$^{20}$ cm$^{-2}$ at 12 $\mu$m and 25 $\mu$m,
respectively. The optical depth per unit H atom in the diffuse ISM is 
$\sigma_{353}=1.2\times 10^{-26}$ cm$^{2}$/H \citep[extrapolating using  $\beta=1.75$ from the 250 $\mu$m value of ][]{boulanger96}. We make a slight correction from the DIRBE to {\it WISE} wavelengths using the NGC 7023 reflection
nebula spectrum, for which 
WISE(12 $\mu$m)/DIRBE(12 $\mu$m)=1.060 and WISE(22 $\mu$m)/DIRBE(25 $\mu$m)=0.92 taking into account 
the spectral responses of
the wide-band filters.
Then, using only these empirical values, the intensity per unit optical depth for the diffuse ISM is
$\epsilon_\nu\equiv I_\nu/\tau_{353}$=2.5 and 2.3 $\times 10^4$ MJy~sr$^{-1}$ for the {\it WISE} 12 and 22 $\mu$m bands, respectively. 
Comparing to Figure~\ref{w3tau}, the clouds often have higher 
$\epsilon_{12}$ and $\epsilon_{22}$ than the diffuse ISM.




{\bfc 
The mid-infrared color ratios of the clouds from the {\it WISE} observations have an average and standard deviation
$\epsilon(12)/\epsilon(22)=0.89\pm 0.24$. For comparison, the model of \citet{lidraine} that matches {\it COBE}/DIRBE
observations of the mid-infrared diffuse interstellar medium has a value 1.69 for that same ratio. 
A wide range of mid-infrared colors was found in a study of somewhat larger
clouds, using {\it IRAS} data, focused on the CO-bright regions,
where \citet{verter00} found $0.5 < F_\nu(12)/F_\nu(25) < 6$ with the cloud fluxes $F_\nu$ in Jy. 
Converting from the {\it IRAS} wavelengths to those of {\it WISE} to get $I_\nu(12)/I_\nu(25)$ reduces those
numbers by only 6\%. Thus it appears our sample of $10^2 M_\odot$ clouds has a fairly narrow range of mid-infrared colors 
that is within the wider range spanned by large clouds and slightly lower than the diffuse interstellar medium. 
The mid-infrared color is sensitive to the relative amount of PAH and very small grains in the dust distribution.
The relatively narrow range of colors in our cloud sample shows that the relative amounts of these two types of
grains appears to remain approximately constant from cloud to cloud.

The mid-infrared emissivity of our $10^2 M_\odot$ sample clouds appears to be significantly higher than for
dark clouds. We showed above that MBM~12 is 2 orders of magnitude lower in emissivity, and that there are large 
variations in the mid-infrared emissivity within that cloud. This seems in keeping with the results
of \citep{verter00} and \citep{bernard93} who showed order-of-magnitude variations in PAH content compared to
larger grains. Our translucent cloud sample has a narrower range of mid-infrared emissivities, which appear to be reasonably 
constant within a cloud, in contrast to the wide range of values and variations within dark clouds. The most
dramatic difference is that PAH appear to be absent from the center of MBM 12.
}

\subsection{Large grain temperatures and radiation field}

We can  address the trend of large grain properties within the cloud sample in a manner similar to that done
in the previous section for the PAH and small grains. 
For the dust temperature, we assume the interstellar radiation field heats the diffuse medium to 20 K (typical
of the diffuse ISM), then the
radiation field decreases from the interstellar average by the same factor $\chi$ as for the PAH and small
grains. In principle the optical depth relevant to the big grain heating can be different from that for 
the PAH, e.g. if the latter were only heated by UV photons. We will take the initial assumption that all
grains are heated by photons with the same characteristic optical depth, specifically that of visible photons.
{\bfc
If the large grain far-infrared emissivity scales as a power-law with exponent $\beta$, equating emission and absorption 
\begin{equation}
\langle\tau_{abs}\rangle_{\rm vis} \chi \propto \langle\tau_{abs}\rangle_{\rm FIR} T ^{4+\beta}
\end{equation}
where the optical depth for absorption, $\tau_{abs}$, in angle brackets, 
is averaged respectively over the visible wavelengths of heating photons and the 
far-infrared wavelengths of cooling photons.
Then we expect the dust temperature
\begin{equation}
T = T_0 \left[ \frac{\rho_0}{\rho}\chi\right]^{\frac{1}{4+\beta}}.
\label{eq:temperature}
\end{equation}
where $T_0$ is the temperature for grains heated by the unextinguished interstellar radiation field at $\chi=1$,
\begin{equation}
\rho\equiv \frac{\langle\tau_{abs}\rangle_{\rm FIR}}{\langle\tau_{abs}\rangle_{\rm vis}},
\end{equation}
and $\rho_0$ is the value of $\rho$ in the diffuse medium where $\chi=1$.

\begin{figure}
\plotone{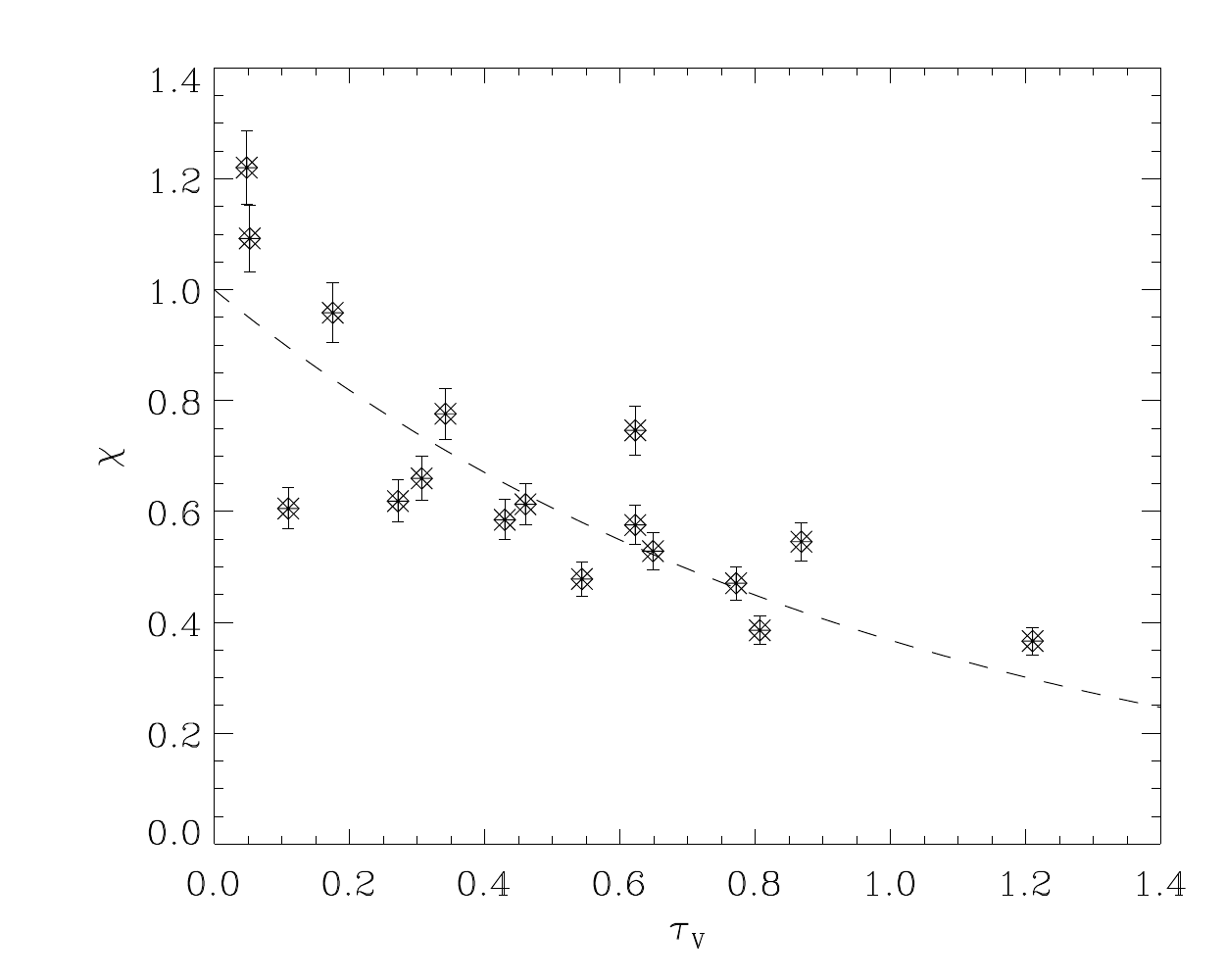}
\caption{The attenuation factor of the radiation field (derived from the big grain dust temperature using eq.~\ref{eq:temperature}) 
optical depth to visible-light photons (derived from the {\it Planck} optical
depth). The dashed line is equation`~\ref{eq:chi}, with no free parameters, for illustration. 
\label{tauchi}}
\end{figure}

If the dust properties are independent of depth into the cloud, then the extinction of the radiation field is determined 
solely by the dust optical depth.
To a crude approximation, the radiation field decreases with depth into the cloud as
\begin{equation}
\chi = e^{-\langle\tau_{ext}\rangle_{\rm vis}} = e^{-\tau_{353}/2.3\times 10^{-5}},
\label{eq:chi}
\end{equation}
where we used  
$\langle\tau_{ext}\rangle_{\rm vis}=4\times 10^4 \tau_{353}$, using the V-band (0.55 $\mu$m) absorption as representative
of visible-light heating photns, in the dust model of \citet{dl07}.
Figure~\ref{tauchi}
shows the attenuation factor $\chi$ (derived from the dust temperature using  eq.~\ref{eq:temperature})
versus cloud optical depth, $\tau_{353}$.
The dashed line is simply eq.~\ref{eq:chi}, which is just the effect of extinction, with the 
radiative transfer treated as absorption, neglecting details of scattering for simplicity.
~
There is significant scatter in Figure~\ref{tauchi} with respect to the simple fit. Some of this 
is due to neglect of radiative transfer (scattering), some due to grain property or 
local interstellar radiation field variations, and some is due to treating the clouds as single 
points rather than taking into account their properties versus depth, which would require a detailed study of each cloud. The `positive' deviants in Figure~\ref{tauchi} have big grains
that are warmer than expected. These include the anomalously warm cloud DIR015+54.
The `negative' deviants include colder clouds, and we verified that all clouds below the dashed line in Figure~\ref{tauchi} have at least some molecular gas detected via millimeter-wave CO emission lines \citep{reach94}.

Could the trend be caused by a change in the properties of the grains? 
Referring to eq.~\ref{eq:temperature}, and assuming for the moment there is no extinction of the radiation field, 
the observed trend of dust temperatures could occur if $\rho$ 
increases as the cloud optical depth increases. Specifically $\rho$ must increase 
by 50\% for clouds with $\tau_{353}=1.1\times 10^{-5}$ (equivalently, $A_{\rm V}=0.5$), and $\rho$ must increase
by 150\% for clouds with $\tau_{353}=2.3\times 10^{-5}$ (equivalently, $A_{\rm V}=1$).
For denser clouds with $A_{\rm V}=2$, $\rho$ must be fully 6 times higher than its diffuse ISM value.
To explain the dust temperatures (Fig.~\ref{tauchi}), either $\langle\tau_{abs}\rangle_{\rm FIR}$ is increasing, 
or $\langle\tau_{abs}\rangle_{\rm vis}$ decreasing, for clouds with greater optical depth.
If $\langle\tau_{abs}\rangle_{\rm FIR}$ is increasing, then we would expect relatively higher far-infrared dust opacities per
unit column density in the higher optical depth clouds, which can be measured by comparing $\sigma_{353}$ to $\tau_{353}$ for
a wide range of clouds.
Figure~\ref{sigtau} shows that there is indeed a correlation such that there is somewhat
more far-infrared optical depth per unit gas
column density for the clouds with higher optical depth. This is another version of the primary effect we explore in this
series of papers, where there is relatively more dust per unit gas for higher-column-density clouds. 
But the apparent increase in FIR opacity is not large enough; 
the changes in dust temperature due to changes in $\langle\tau_{abs}\rangle_{\rm FIR}$ appear to cause 
only a small part of apparent excess dust. 
}

\begin{figure}
\plotone{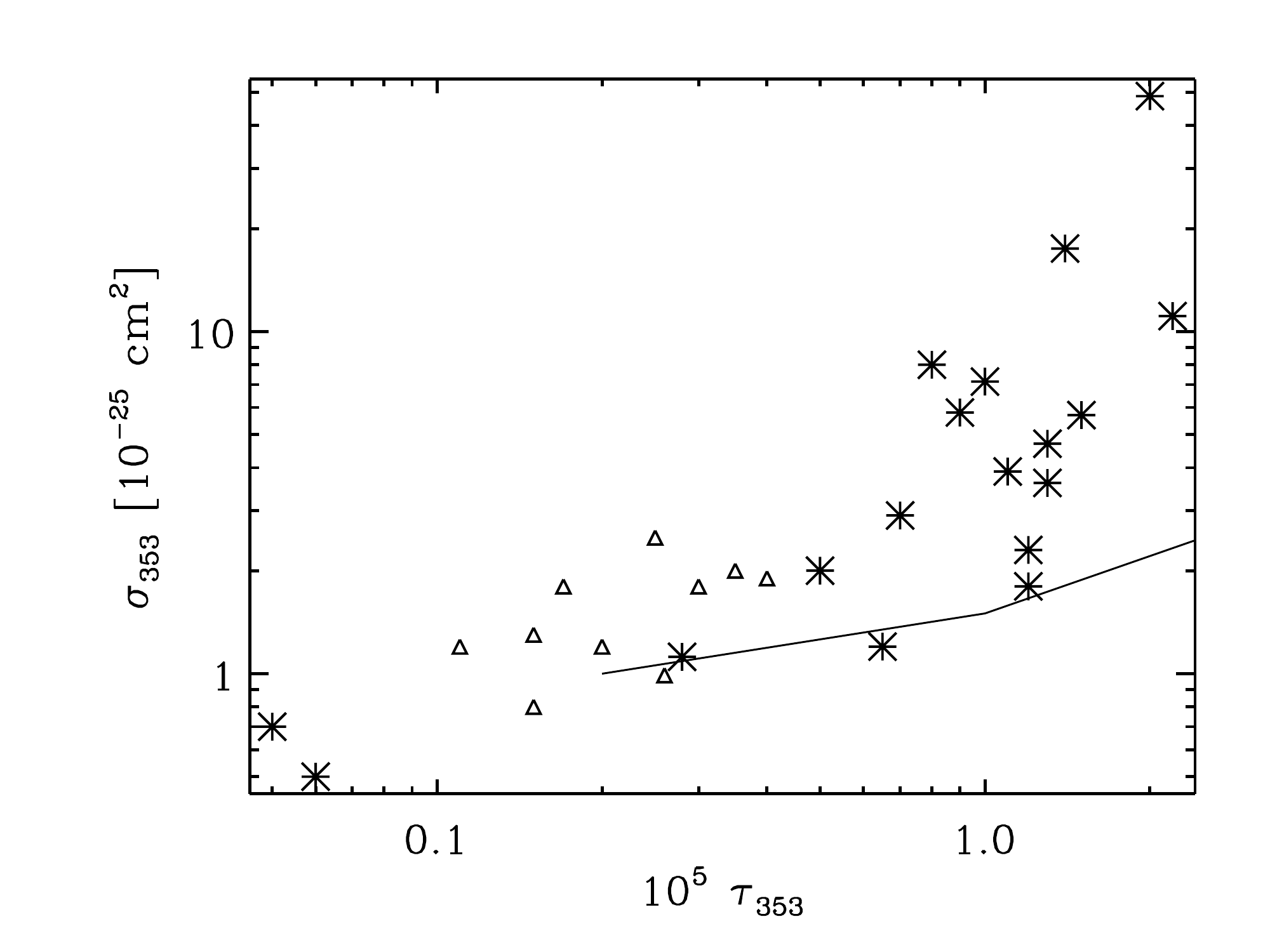}
\caption{The far-infrared dust optical depth per unit gas column density, emissivity $\sigma_{353}$ versus the cloud
optical depth for our sample clouds. Asterisks are for the cloud peaks, and open triangles are for the background regions of the
clouds in the Arecibo survey. The solid line shows the trend that is predicted as a reflex of the correlation between dust temperature and
optical depth. 
\label{sigtau}}
\end{figure}

If we were to ascribe all the grain temperature variation to changes in $\langle\tau_{abs}\rangle_{\rm vis}$, then
optical and ultraviolet measurements of starlight extinction would show corresponding trends
in extinction per unit column density. But $A_{\rm V}/N_{\rm H}$ has been well measured from starlight
extinction and ultraviolet H column densities. For example, in the {\it FUSE} and {\it Copernicus} survey reported by \citet{rachford02},
there is a linear correlation between extinction and the total column density, $N_{\rm H}+2 N_{{\rm H}_2}$ over a range
of extinctions from 0.15 to 3.2, corresponding to $\tau_{353}$ from $4\times 10^{-6}$ to $8\times 10^{-5}$. 
That means that the optical depth to absorption of starlight $\langle\tau_{abs}\rangle_{\rm vis}$ per unit gas column density
does not change, and it cannot explain the variation in dust temperatures.
We conclude that the big grain temperature changes in our clouds are largely due to changes in the radiation
field, not due to dust property variations.

\subsection{Direct comparison between PAH and big grain abundance}

The emission from PAH is via excitation by single photons, followed by cascade through
all the infrared modes, so the mid-infrared brightness is proportional to the UV-visible radiation field strength and the column of PAH:
\begin{equation}
I(12) \propto \chi Y_{\rm PAH} N_{\rm H}
\end{equation}
where $Y_{\rm PAH}$ is the abundance of PAH.
The true gas-to-dust ratio, which is dominated by that of large grains, should be uniform in the ISM because most of the heavy-element
abundance is locked in grains and cannot vary significantly from place to place. Thus we assume the
abundance of large grains $Y_{\rm BG}$ is constant, and the dust optical depth traces the total column.
In that case, the emissivity of PAH depends on the radiation field and PAH abundance:
\begin{equation}
\epsilon(12) = I(12)/\tau_{353} \propto \chi Y_{\rm PAH}.
\end{equation}
Assuming the large grain properties are uniform, we can use the observed the dust temperature to 
measure $\chi$ using eq.~\ref{eq:temperature}
with $\rho=\rho_0$ and $T_0=20$~K.

\begin{figure}
\plotone{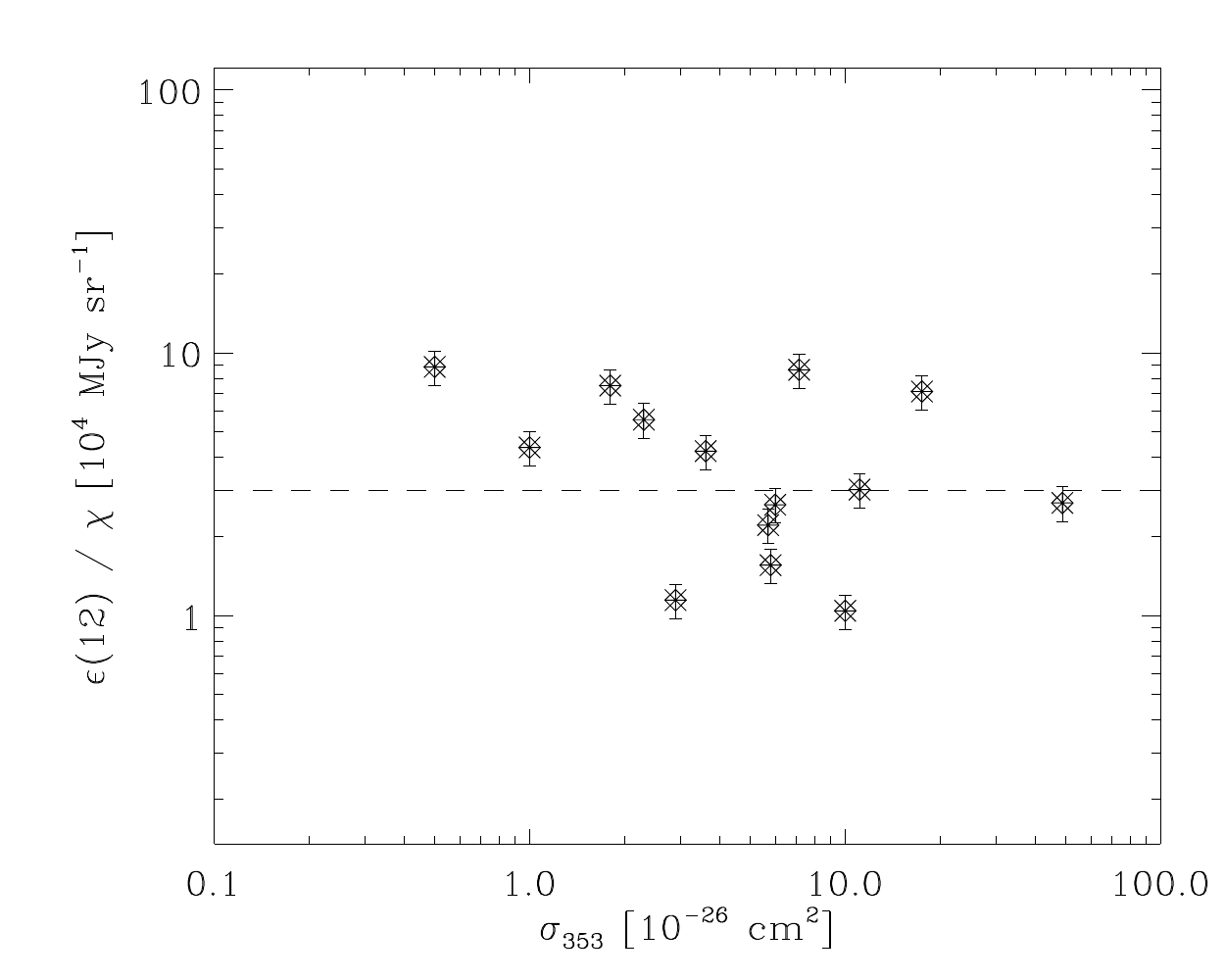}
\caption{Ratio of PAH to big grain abundance versus abundance of big grains for our cloud sample. 
The dashed line is a one-to-one correlation for illustration. 
\label{bgpah3}}
\end{figure}

Figure~\ref{bgpah3} directly compares the observables
\begin{equation}
\epsilon(12)/\chi \propto Y_{\rm PAH}  
\end{equation}
versus
$\sigma_{353}$
The observations are consistent with a constant value, shown by the dashed line, showing there is little room
for variation in the abundance of PAH. Using the standard deviation of $\epsilon(12)/\chi$, we find
that $Y_{\rm PAH}$ remains constant within 60\%.
The results for the abundance of very small grains, traced by the {\it WISE} 22 $\mu$m emission, are identical, 
showing $Y_{\rm VSG}$ is also constant to within 60\%.
Thus we can accept a null hypothesis that the relative abundance of PAH and very small grains remains approximately constant within 
the clouds, and the apparent decrease in brightness
is due to extinction of the radiation field.

{\bfc
\subsection{Transition from diffuse to dark clouds}

Our study applies to local $10^2 M_\odot$ clouds and takes advantage of {\it WISE} and {\it Planck} 
observations that clearly resolve them.
A study of  nearby, dense clouds with bright CO emission showed some have halos of mid-infrared emission that are
explained by increases, by factors of few, in PAH and very small grain abundances in some cloud halos, even when
taking into account radiative transfer of starlight into clouds
\citep{bernard93}. Not all clouds show the strong mid-infrared limb-brightening of the Chamaeleon and Ophiuchus clouds
studied by \citet{bernard93}, so those strong variations in small grain abundances may be local effects related to
the cloud history of dynamic interactions with other clouds and nearby stars.
Another study of infrared emission of larger, CO-bright clouds with cold, dense cores 
showed that their colors may be more influenced by grain coagulation than the extinction of starlight
 \citep{ysard12}.

To assess the total amount of gas in dark and molecular clouds, it is customary to use the brightness of the
conveniently-observed lowest rotational transitions of CO, then convert to the H$_2$ column density with 
a scale factor \citep{xfactorreview}. That factor has some theoretical basis if the spectral lines are optically thick,
the gas temperature is constant, and the clouds are self-gravitating so the velocity dispersion determines the mass
via the virial theorem. For clouds that are on the border between being diffuse and self-gravitating, these assumptions
do not apply, leading to the possibility of significant molecular gas without CO \citep{blitz90}. Observations of
the small-scale structure of CO emission in a high-latitude molecular cloud show that the velocity structure and 
spatial distribution of CO in fact have nothing to do with the virial theorem or distribution of H$_2$, being instead
due to intermittent turbulence \citep{falgarone09}.
We require a tracer of the total column density that can operate in regions with column densities up to $10^{22}$ cm$^{-1}$
(visible extinctions of a few). One such tracer is radio CH emission \citep{magnani03}, though that emission is faint and
cannot be readily detected and mapped for diffuse clouds.
The interaction of cosmic rays with nucleons gives rise to $\gamma$-rays, which can then trace the total column density of
interstellar nucleons. Both the cosmic rays and $\gamma$-rays can propagate through long distances in the Galaxy.
It is notable that the usage of $\gamma$-rays is completely independent of the dust and gas measurements, and it serves as
a test of the capability of dust to be used as a tracer of the total column density.
By comparing the $\gamma$-ray derived total column density to that the gas tracers, it was possible to separate the gas into
the part that can be traced and the part that is inferred to exist but is `dark' \citep{grenier05}.
A {\it Fermi} study of dark clouds in the galactic anticenter 
measured $\sigma_{353}=1.6\times 10{-26}$ cm$^2$ using $\gamma$-rays specifically in the `dark' gas \citep{remy17}, consistent with
Figure~1 for diffuse gas.
\citet{remy17} found a significant increase in $\sigma_{353}$, by a factor of 6 in dark gas with $T_{\rm d}<14$~K, in locations with dense gas, indicating grain property evolution.
The direction and amount of the grain property variations is in agreement with extrapolation of the trends seen in the diffuse medium and
shown in Figure~1.  Whether dust emission can be used as a reliable tracer of
total column density in dense clouds is in question, because it is not known how repeatable and systematic are the changes in large
grain dust properties that affect $\sigma_{353}$. 
For diffuse clouds, with temperatures greater than $T_{\rm d} \ge 16$~K, and where CO emission is not bright,
we find that the change in dust properties is still small enough that
the total column density can be determined from the dust emission.

In Figure~\ref{dgvarfig}, the solid points indicate the dust absorption per unit atomic gas column density, and they show
a clear increase for clouds that are colder. For some clouds, we have CO observations and can use the conventional
method of multiplying the CO line integral, $W({\rm CO})$ 
by the factor $X\equiv N({\rm H}_2)/W({\rm CO})=2\times 10^{20}$ cm$^{-2}$~K$^{-1}$~s, to estimate the molecular column density. 
Then we can calculate the dust absorption per unit {\it total} (atomic plus molecular)
gas column density; those values are shown as open symbols 
for 4 clouds in Figure~\ref{dgvarfig}. 
If the dust properties in atomic and molecular clouds are identical and invariant,
and we have correctly estimated the total gas column density, then the open symbols should fall on the same line as
the atomic clouds.
For the three diffuse clouds from our main sample, the open symbols do in fact fall on the line that was derived for
diffuse atomic clouds.
It therefore appears that, taking into account
atomic and molecular gas, the true value is $\sigma_{353}^{tot}\simeq 1.5\times 10^{-26}$~cm$^{2}$ for clouds wuth
dust temperature of 18~K.
The gradual slope of the solid line in Figure~\ref{dgvarfig} represents a change in dust
properties, being a factor 
of 2 higher for colder clouds (down to dust temperatures of 15 K) and factor of 2 lower for warmer clouds 
(up to dust temperatures of 20.5 K).

The large, vertical deviations of the clouds in Figure~\ref{dgvarfig} from the solid line can be interpreted
as due to unaccounted molecular gas, or to anomalous grain properties in the clouds. The systematic trend,
the agreement with the model for H$_2$ formation, suggest the primary effect is presence of molecular gas.
Note that the H$_2$ formation model depends on the volume density of the gas and really requires an
analysis that goes cloud-by-cloud and has a realistic radial density profile connecting to the diffuse ISM.
For the dark cloud MBM~12, taking into account the molecular gas traced by CO 
does not bring its $\sigma_{353}^{tot}$ quite in line with the extrapolation of dust properties from 
atomic clouds. The dust absorption per unit column density appears to be another factor of 2 higher than the 
linear extrapolation of the dashed line 
in Figure~\ref{dgvarfig}. This could indicate that dust property variations become even more
important in molecular clouds than in diffuse clouds, in the same sense and approximate magnitude
as determined from $\gamma$-ray studies \citep{remy17}.
}

\section{Conclusions}

{\bfc 
Dust grain properties vary in diffuse interstellar clouds, but that variation is insufficient to explain the observed range
of dust emission per unit atomic gas column density. 
Within our sample of isolated, diffuse clouds
the relative amounts of PAH, Very Small Grains, and Big Grains remain approximately constant to within a factor of 2. 
The  trends in the relative amount of emission from different types of grain, and he large grain temperature differences,
can be explained
by a decrease of the radiation field from due to extinction of the starlight
that heats the grains.
In the core of a high-latitude dark cloud, there is no evidence for any PAH at all, likely due to the absence of heating photons
but also consistent with absence of PAH themselves.

In Paper 1, we showed significant variations in the gas-to-dust ratio, when the gas is traced by the 21-cm line of \ion{H}{1}.
In Paper 2, we showed the H~I was not significantly underestimating the atomic column density, because there is little column
density of 21-cm-optically-thick, cold atomic gas. 
In this Paper, we find no evidence that changes in dust properties that can explain the apparently high gas-to-dust ratio
(equivalently, higher specific opacity) of diffuse clouds. 
While there is evidence for a modest increase in the far-infrared opacity in the diffuse interstellar medium, 
the total range of observed gas-to-dust ratios is too large
to be explained by this effect alone.
An updated empirical model for H$_2$ formation on large grain surfaces falls far short of the atomic gas-to-dust ratio, with extrapolation
to low radiation fields yielding predictions consistent with a high-latitude dark cloud known to contain significant molecular gas.
In terms of the hypotheses posed in the Introduction, we show that while there is a small amount of cold atomic gas (hypothesis 2) and
there are modest dust property variations (hypothesis 3),
the results support that diffuse clouds contain `dark gas' (hypothesis 1), presumably H$_2$, not traced by the \ion{H}{1}, 
even in places with little CO emission. Where there is CO emission, the dust optical depth per unit total gas is consistent with the
trend of dust optical depth per unit atomic gas in diffuse clouds.
In dark clouds with extinction $A_{\rm V}>2$, where there is bright CO emission and the dust grains are cold 
($T_{\rm d}<16$ K), dust property variations may be more extreme than in diffuse clouds.
}

\acknowledgements  
This publication makes use of data products from the Wide-field Infrared Survey Explorer, which is a joint project of the University of California, Los Angeles, and the Jet Propulsion Laboratory/California Institute of Technology, funded by the National Aeronautics and Space Administration.
Facilities: \facility{Planck}, 
\facility{WISE}

\bibliographystyle{aasjournal}
\bibliography{wtrbib}

\end{document}